\documentclass[final,english]{bullsrsl}[2022/06/15]

\usepackage[latin1]{inputenc}
\usepackage[T1]{fontenc}

\usepackage{natbib} 
\usepackage{graphicx}
\usepackage{pdflscape}

\begin{document}
\title{Photometric Variability of Low Mass Stars and Brown Dwarfs in IC~348 and Taurus Star-Forming Regions}

\author[affil={1},corresponding]{Samrat}{Ghosh}
\author[affil={1}, corresponding]{Soumen}{Mondal}
\author[affil={2}]{Santosh}{Joshi}
\author[affil={2}]{Sneh}{Lata}
\author[affil={1}]{Rajib}{Kumbhakar}

\affiliation[1]{S. N. Bose National Centre for Basic Sciences, Salt Lake, Kolkata 700106, India}
\affiliation[2]{Aryabhatta Research Institute of Observational Sciences (ARIES), Manora Peak, Nainital-263002, India}
\correspondance{samrat687@gmail.com; soumen.mondal@bose.res.in}
\date{24th May 2023}
\maketitle


%

\begin{abstract}
Low-mass stars belonging to the M spectral type are the most numerous stars in our Galaxy, amounting to about two-thirds in number, and are found at the bottom of the main sequence in the H-R diagram. 
Photometric studies of low-mass stars, including brown dwarfs (BDs), provide several important evolutions of their atmosphere, magnetic flares and chromospheric activity. This paper highlights a few interesting results from our optical I-band observations of 2MASS J03435638+3209591 in the young star-forming IC~348 region and three BDs in Taurus star-forming regions using ground-based telescopes as well as a space-based telescope. We estimated the fast periodicities in the range of 1.5 to 3 hours in Taurus BDs. Furthermore, using the long-term photometry from the Transiting Exoplanet Survey Satellite (TESS), we have conducted a time-resolved variability analysis of CFHT-BD-Tau 4. The periodogram analysis of TESS data reveals an orbital period of $\sim$ 3 days. We found two flare events in TESS sector 43 data for this BD and estimated the flared energies as $4.59\times10^{35}$ erg and $2.64\times10^{36}$ erg, which sit in the superflare range. 
\end{abstract}


\section{Introduction}
Very low-mass stars (VLMs) refer to the stellar and substellar objects with masses below 0.6 $M_\odot$ to planetary limit (0.013 $M_\odot$) \citep{1997ARA&A..35..137A}, which include the spectral type of mid-K, M, L, T to the coldest known brown dwarf Y \citep{2012RSPTA.370.2765A}. They extend from the edge of the hydrogen-burning main sequence (MS, i.e., mid-K and early-M) to the deuterium-burning Brown Dwarfs (BDs; mass-range 80 - 13 $M_J$) with the transition at M6 spectral type at the young age of star-forming regions (SFRs) \citep{2000ARA&A..38..337C, 2016ApJ...827...52L}. Choosing a star-forming region seems technically suitable for studying these objects with ground-based telescopes, as their luminosities and temperatures are higher than galactic field BDs and low-mass stars. Photometric variability studies can detect periodic and aperiodic variability in young BDs and probe the nature of their atmospheres \citep{2014ApJ...796..129C}. Stable surface features like star spots or dust clouds cause non-uniformity in the surface, which causes periodic optical modulation of the flux as the object rotates. Whereas aperiodic variability is due to non-uniform accretion from disks, magnetic spots, and dynamic star spots or dust clouds, changing with time scales different from the rotation periods. VLMS, including BDs, are fast rotators with periods ranging from a few hours to days \citep{2010ApJS..191..389C, 2014yCat..35660130C}. The average rotation rate of stars varies across the main sequence, differing in stellar structure and magnetic properties. The periodic variability can be observed within a few nights of photometric monitoring using 1-2 m class ground-based telescopes. 

We highlight here the findings of the optical I-band photometric variability of VLMs and BDs from two star-forming regions, IC~348 and Taurus. IC 348 is a young (1 - 3 Myr; and nearby $\sim 310$ pc; \citealp{2003ApJ...593.1093L}) star-forming region in Perseus molecular cloud. Because of IC 348 cluster's intermediate star density ($ \rho \sim 100-500~M_\odot ~pc^{-3}$; \citealp{2017MNRAS.468.4340P}), it has enough stars ($\sim$500) to detect large populations of low-mass objects in 10-20 arcmin field of view (FoV), and similarly for Taurus. Our optical {\it $I$}-band photometric variability study down to $\simeq$ 19 magnitudes in IC~348 explored the hour-scale rotation in VLMs. We present the findings of a young M dwarf 2MASS J03435638+3209591 (SpT= M7.25 V, I=18.61) in IC 348. This BD is reported to be a possible variable source \citep{ 2014ApJ...796..129C, 2017ApJS..229...28G}. BDs are generally fast rotators. This makes this object an object of interest. Moreover, deuterium burning instability or $\epsilon$-mechanism is induced in the core of fully convective VLMs and BDs \citep{2005A&A...432L..57P} due to the high sensitivity of nuclear energy generation rate to temperature ($\sim T^{12}$), which could give a pulsation period that varies between 1h to 5h, which in turn could be observed as a short periodicity in the data with observable amplitude.

We have monitored three known brown dwarfs: CFHT-BD-Tau 2 (hereafter, CT2), CFHT-BD-Tau 3 (hereafter, CT3), and CFHT-BD-Tau 4 (hereafter, CT4) (Age $\sim$ 1 Myr; \citealp{2001ApJ...561L.195M}) in Taurus Molecular clouds with different Indian national telescope facilities on various epochs. BD nature of CT2, CT3, and CT4 has been confirmed spectroscopically \citep{2001ApJ...561L.195M}. We also conducted a time-resolved variability analysis of CT4 using the 2-min cadence data from the Transiting Exoplanet Survey Satellite (TESS). Previous studies revealed that CT2, CT3, and CT4 have day-scale rotation periods of 2.93, 0.96, and 2.95 days, respectively \citep{Scholz_2018, 2020AJ....159..273R} in K2 mission observations.Magnetic activity is common in M dwarfs, which is supported by the flaring of CT4 and previously reported $H_{\alpha}$ detection \citep{2005ApJ...626..498M} for this BD. Previous studies have also shown that CT4 is accreting, but CT2 and CT3 are non-accretors \citep{2005ApJ...626..498M}.

\section{Observations and Data Analysis}
\label{obs_sect}
The $I$-band optical photometric data for 2MASS J03435638+3209591 were obtained from the 1.3-m Devasthal Fast Optical Telescope (hereafter, 1.3m DFOT) operated by ARIES, Nainital and 2-m Himalayan Chandra Telescope (hereafter, 2-m HCT), Ladakh operated by the Indian Institute of Astrophysics (IIA), Bangalore. Data for Taurus BDs were acquired using the 1-m Sampurnanda Telescope (hereafter, 1-m ST) and 1.3-m DFOT operated by ARIES, Nainital, and the 2-m HCT. The Wright $2K$ CCD with 24$~\mu m$ pixel size is used in 1-m ST for acquiring the data with a field of view (FoV) of 13$\times$13 $arcmin^2$ (\citealt{2014PINSA..80..759S}). The ANDOR $2K\times2K$ CCD in 1.3-m DFOT has a pixel size of $13.5 \times 13.5 ~\mu m^2$. A set of Johnson-Cousin $B$, $V$, $R$, $I$, and $H_\alpha$ circular filters are available which gives an unvignetted view of $18\times18 ~arcmin^2$. The back-end instrument used in 2-m HCT is the Himalayan Faint Object Spectrograph and Camera (HFOSC). The $2K \times 2K$ part of the detector in $2K \times 4K$ CCD having a pixel size of 15 $\mu m$ and a pixel scale of 0.296 arcsecs is used for imaging observations. The FoV on the $2K \times 2K$ part of CCD in the imaging mode is $10 \times 10$ $arcmin^2$. In all these instruments, photometric images were taken in the optical $I$-band with exposure times varying according to the night condition (see Table\,\ref{log-table}). 

Optical I-band data reduction was performed using Image Reduction and Analysis Facility (IRAF) software. Bias, flat, and cosmic ray corrections were performed to each observed frame using standard IRAF tasks. Aperture photometry using {\sc phot} was performed on bias and flat-corrected images to get instrumental magnitudes ($I$-band) of individual sources. We used the estimated instrumental magnitudes for each observing date to construct the time-series light curves for all sources. Following \citet{2021MNRAS.500.5106G}, we generated the light curves using differential photometry for all sources in the CCD frames. We used the Lomb Scargle method (LS periodogram; \citealt{1976Ap...SS..39..447L}; \citealt{1982ApJ...263..835S}, \citealt{2018ApJS..236...16V}) to find the significant periodicity of the data. Phase light curves have been constructed using the most significant peak. We used the Python package {\it `astropy'} \citep{astropy:2013} to bin the data.

Using our new I-band observations from a sample of 177 light curves of IC~348, we detected 22 young M-dwarfs, including 6 BDs. Out of 22 variables, $\sim$50 per cent show hour-scale periodic variability in the period range of 3.5 - 11 hours; the rest are aperiodic \citep{2021MNRAS.500.5106G}. Here, we presented one BD source as an example of the short rotation periodicity of a faint BD detected with ground-based observation.

Using ground-based observations' time window, we monitored three Taurus BD sources in two different framings (one framing for CT2 and CT3; another for CT4) as they are separated by a larger distance than the FoV of the attached CCD camera on the telescope. We detected and analysed $\sim$200 sources for 1-m ST and 2-m HCT images for CT2 and CT3 frames and $\sim$30 objects for CT4 frames, as this region is less crowded. In 1.3-m DFOT, we detected $\sim$700 sources in 18$\times$18 $arcmin^2$ FoV for the CFHT 2 and CFHT 3 frame. We selected ten sources of similar magnitude from these objects' light curves. We averaged them to create an average reference light curve for each night to generate a differential light curve for the target objects and references.
 
\subsection*{TESS Observations and Data Analysis}
Transiting Exoplanet Survey Satellite (TESS; \citealp{2015JATIS...1a4003R}) observed CT4 with camera 2 during sector 43 from September 16, 2021, to October 11, 2021. The four cameras of TESS as back-end instruments cover the field of view of each $24\times24 ~deg^2$ and are aligned to cover $24 \times 90 ~deg^2$ of the sky, which is called `sectors' \citep{2015JATIS...1a4003R}. The data were stored under the Mikulski Archive for Space Telescopes (MAST) with identification number `TIC 150058662' (TIC: TESS input catalog). We retrieved TESS data with the TIC of the object CT4 from the MAST and processed the light curves from TESS 2-min cadence data using the `lightkurve' package \citep{2018ascl.soft12013L}. We used Pre-Search Data Conditioning (PDCSAP) light curves because these are already corrected for systematic instrumental noise in the Simple Aperture Photometry (SAP) light curves. PDCSAP light curve shows short time-scale flux variation and less scatter \citep{Smith_2012, Stumpe_2014}.

\begin{figure}[t]
\centering
\includegraphics[width=0.75\textwidth]{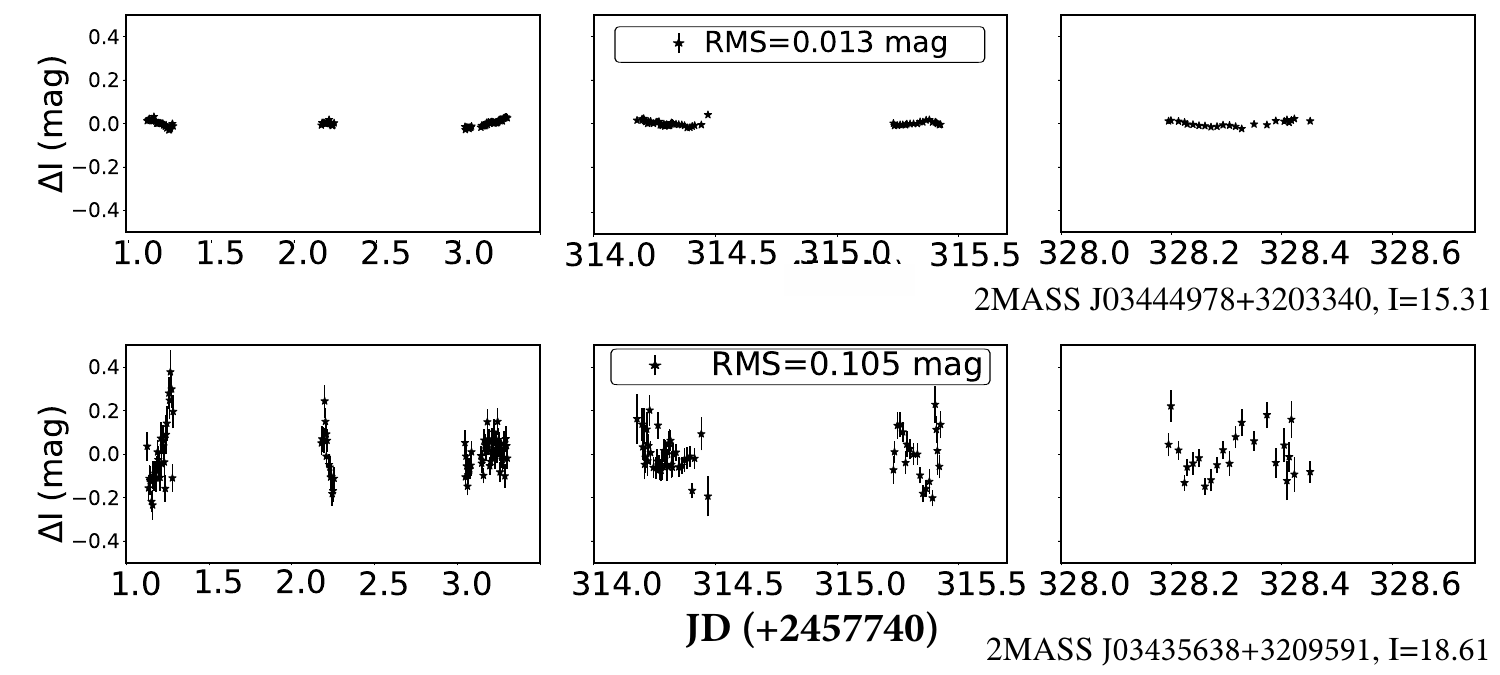}
\smallskip

\begin{minipage}{12cm}
\caption{For example, a light curve of variable 2MASS J03435638+3209591 in the bottom left panel is shown. A non-variable reference star is shown in the top left panel. The x-axis is broken due to data gaps between observations. The y-axis is the zero-averaged $I$ magnitude of the sources. Plot regenerated from \cite{2021MNRAS.500.5106G} with permission.}
\label{fig:1}
\end{minipage}
\end{figure}

\begin{figure}[t]
\centering
\includegraphics[width=0.5\textwidth]{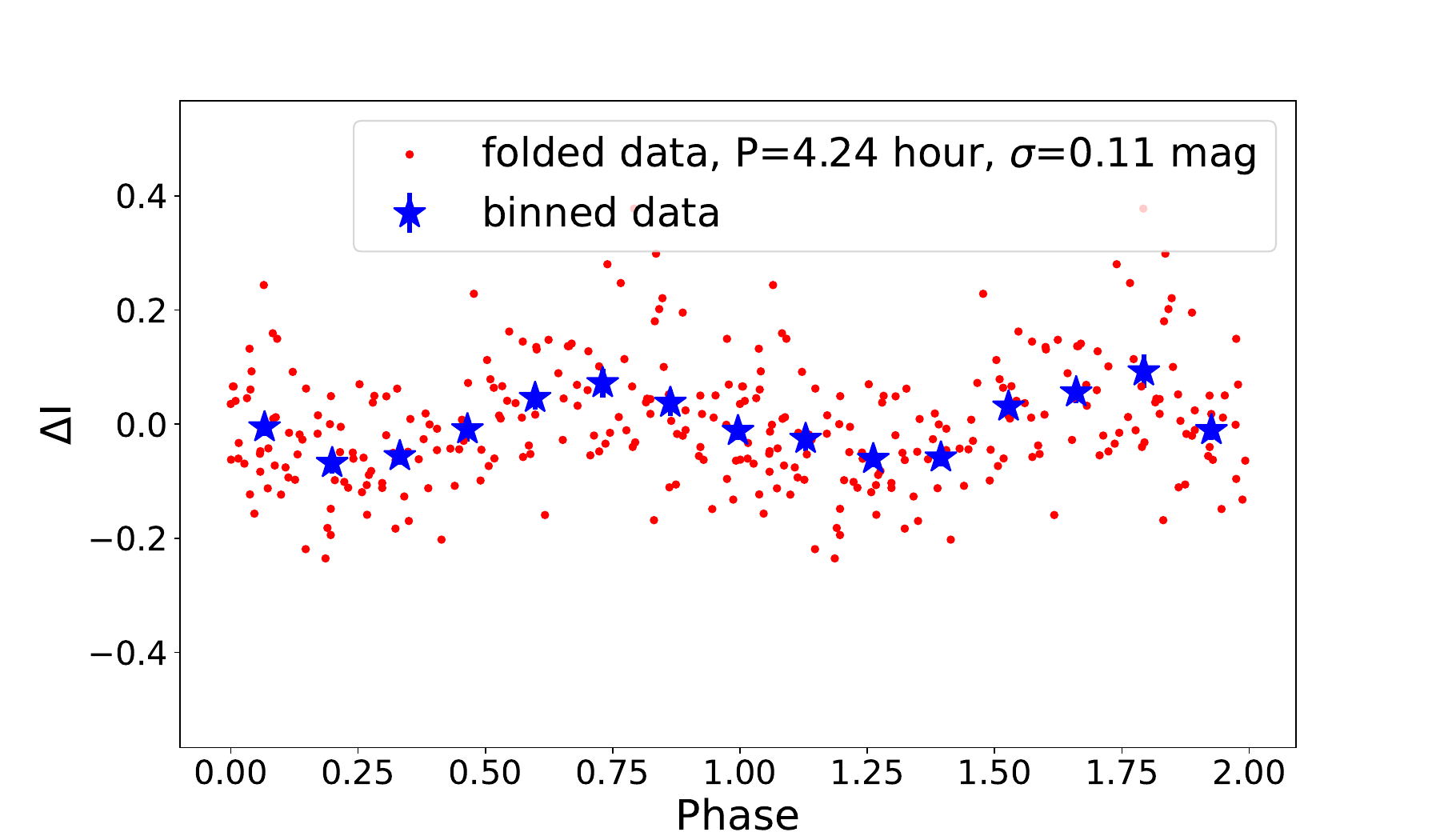}
\smallskip

\begin{minipage}{12cm}
\caption{The phase light curve of the same periodic source with the period, P= 4.24 h in red dots, is shown. The blue stars represent the 15-point binned data. Plot regenerated from \cite{2021MNRAS.500.5106G} with permission.}
\label{fig:2}
\end{minipage}
\end{figure}

\section{Results and Discussion}

The light curve of the variable object 2MASS J03435638+3209591 from IC~348 is shown in
Fig.\,\ref{fig:1},
showing significant variability with RMS=0.105 mag. The light curve of a non-variable source 2MASS J03444978+3203340 (I=15.31) is shown in the top panel with RMS=0.013 mag indicating the object is variable. We found a short period of 4.24 hours from the LS periodogram analysis. The phase folded light curve is shown in
Fig.\,\ref{fig:2}.

\begin{figure}[t]
\centering
\includegraphics[width=\textwidth]{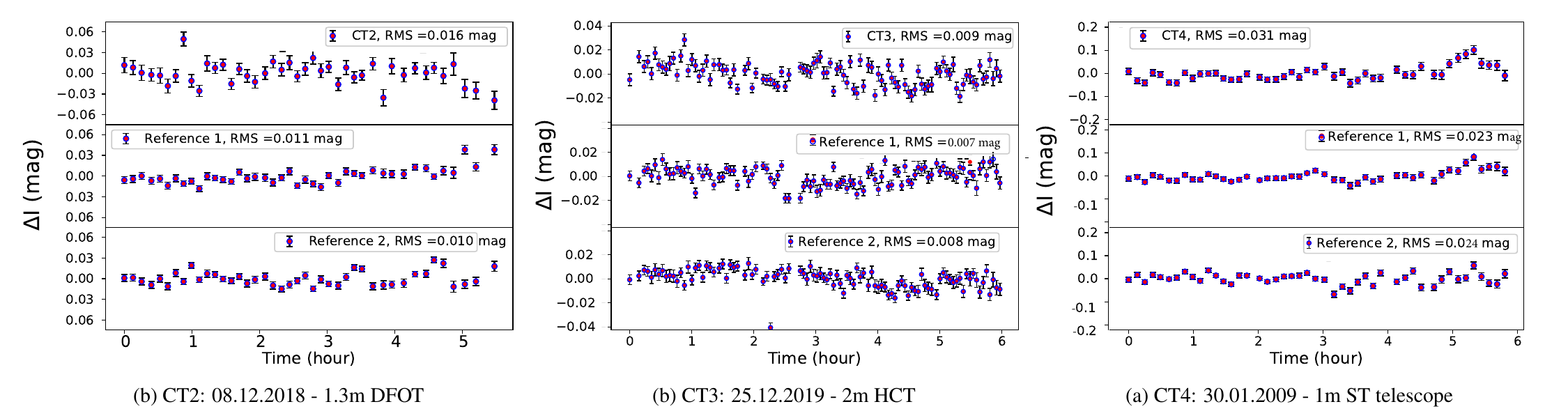}

\begin{minipage}{12cm}
\caption{Three light curves of the three brown dwarfs are shown. Two non-variable reference light curves are shown in the bottom panels of each light curve. The source name, dates, and telescopes used to obtain the data are mentioned underneath each panel.}
\label{fig:3}
\end{minipage}
\end{figure}

\begin{figure}[t]
\centering
\includegraphics[width=\textwidth]{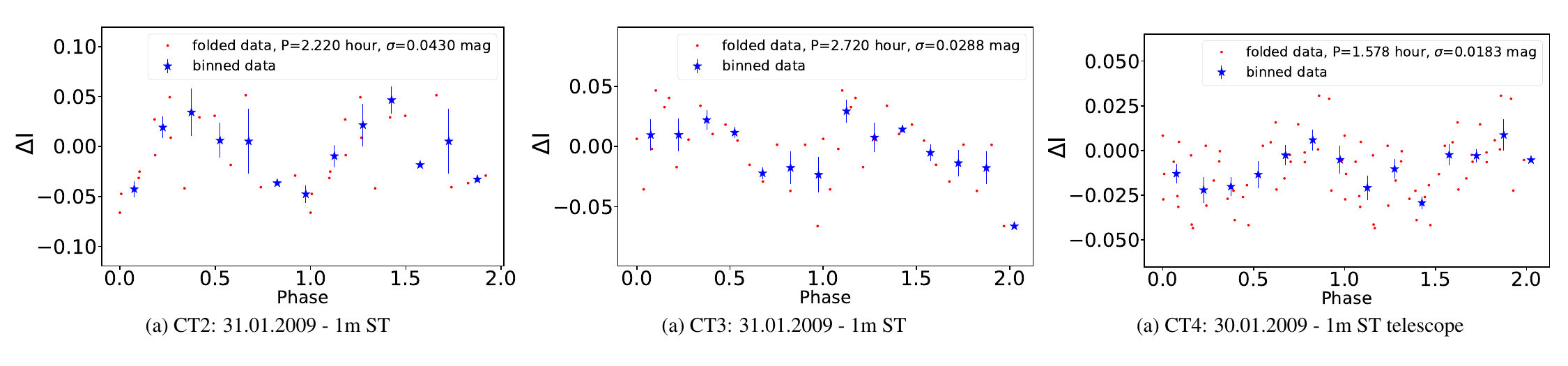}

\begin{minipage}{12cm}
\caption{A few example phase light curves of the BDs are shown. The blue stars represent 10-point binned data binned in phase. The source name, dates, and telescopes used to obtain the data are mentioned underneath each panel.}
\label{fig:4}
\end{minipage}
\end{figure}

Three light curves from three different observing nights are shown in
Fig.\,\ref{fig:3}
for the Taurus BDs. A few phase light curves are shown in
Fig.\,\ref{fig:4}.
Using the LS periodogram method, we uncovered very short hour scale periodicities for all three sources ($\sim$1.5 - 3 h) from the ground-based $I$-band data.


\begin{figure}[t]
\centering
\includegraphics[width=\textwidth]{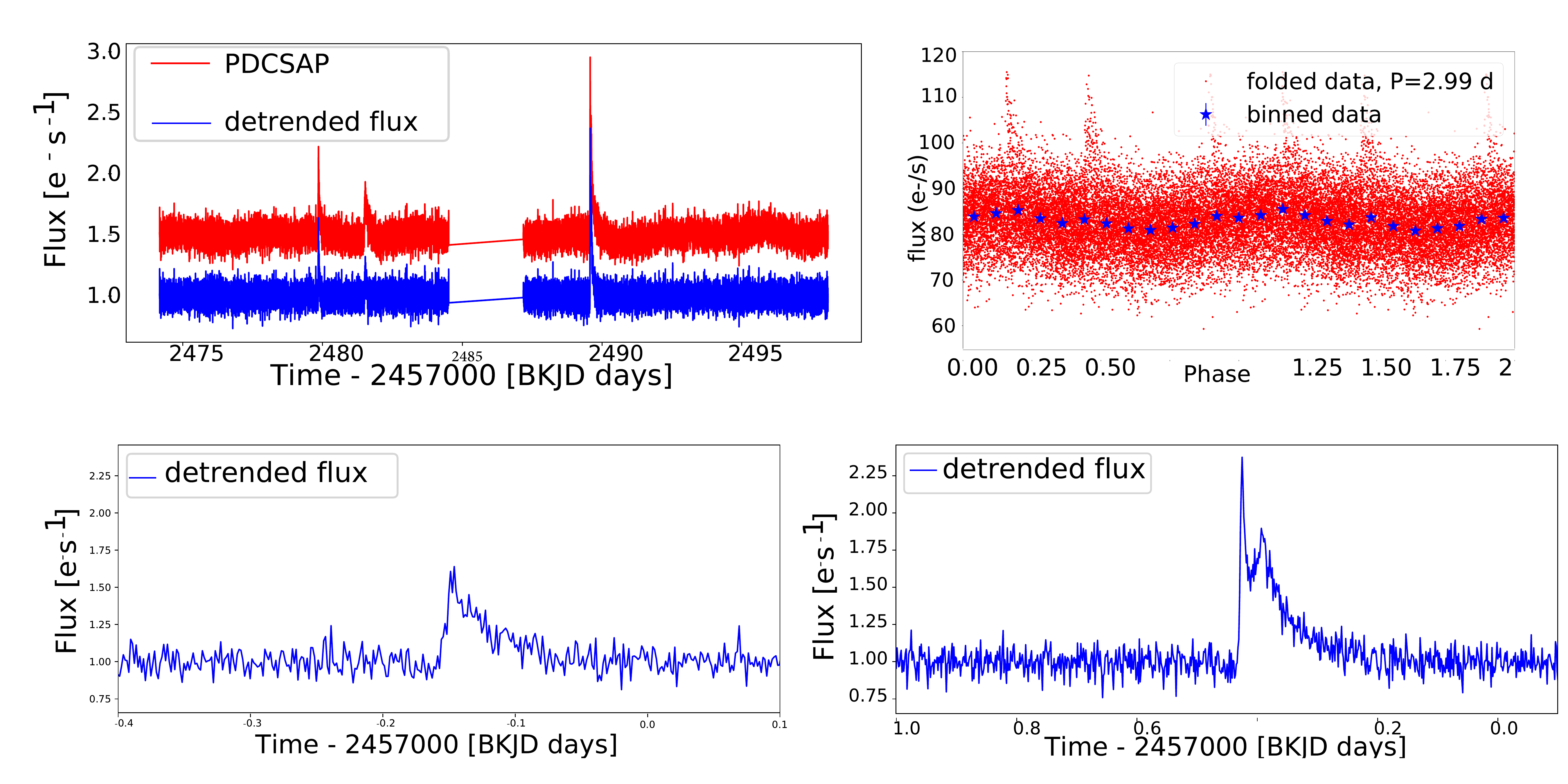}

\begin{minipage}{12cm}
\caption{Upper left panel shows Corrected light curves of CFHT-BD-Tau 4 from sector 43 of TIC 150058662 2-min cadence data. The x-axis is the time in Barycentric Kepler Julian Days (BKJD), and the y-axis is the normalized TESS flux ($e^{-1} /s$). Red highlights PDCSAP data. Blue gives the detrended flux ($e^{-1} /s$) with no intrinsic variation present.  Upper right panel shows the phase light curve of CFHT-BD-Tau 4 folded with the period, P= 2.99 day. The red dots show the phase curve; the blue stars are binned data. The x-axis shows phase and the y-axis shows flux in ($e^{-1} /s$). The bottom panels show zoomed-in view of the two flares.}
\label{tess_ct4}
\end{minipage}
\end{figure}


Using the long-term photometry from the TESS, we have conducted a time-resolved variability analysis of CT4. The LS periodogram analysis of TESS data reveals an orbital period of $\sim$2.99 days, consistent with the earlier results from the literature (2.93 days: \citealt{Scholz_2018} and 2.95 d: \citealt{2020AJ....159..273R}). The PDCSAP and normalized light curves are shown in the top left panel of Fig.\,\ref{tess_ct4}, and the phase light curve, folded with 2.99 days, is shown in the top right panel. The red dots show the phase curve; the blue stars are binned data.\\
The possible explanation of the hour-scale periodicity from ground-based $I$-band data could be accretion hot spots in those BDs, co-rotating with the objects where accretion is still ongoing \citep{refId0} and/or slowly evolving magnetic spots that create such light-curve morphology over a stable rotation period of $\sim 3$ days. Variable extinction from circumstellar disk can also result in either periodic or long timescale variability \citep{2014ApJS..211....3P}. Variable extinction might not always be sinusoidal, unlike variability caused by star spots, but might appear more likely as eclipse-like features \citep{2014ApJS..211....3P}. The disc's rotational speed determines extinction variations time scales, and depending on the radial distance from the star, it could range from a few hours to years \citep{2009MNRAS.398..873S}. If we assign the day-scale periods of CT2, CT3 and CT4 as rotational periods or due to the presence of surface spots; then another explanation of hour-scale variability could be the deuterium burning instability \citep{2005A&A...432L..57P}.  In the mass range of 0.2 to 2.0 $M_{\odot}$, PMS stars could become pulsationally unstable during deuterium burning, which could affect the radiation and make them variable \citep{1972A&A....19...76T}. This instability is induced in the core of fully convective BDs due to the high sensitivity of nuclear energy generation rate to temperature ($\sim T^{12}$). A small temperature perturbation could induce an order of magnitude higher energy variations causing a pulsation period between $\sim$1h to $\sim$5h  and \citet{2005A&A...432L..57P}, which in turn could be observed as a short periodicity in the data with observable amplitude.

Combining this with the fact that the sources have reported day-scale periods, we could infer that the noise from ground-based monitoring may be high enough to mimic this as a periodic signal in short-time scales (hours-scale). These hour-scale periods might give some insights into variability due to little-known physics, even at low significance levels. However, the readers should use these parameters with caution.

We analyzed the TESS sector 43 light-curve of CT4  using the {\it `altaipony'}  pipeline \citep{Ilin2021JOSS....6.2845I}. Altaipony uses {\it FlareLightCurve.find\_flares()} to detect the flare candidates. It uses the detection criteria described in Chang et al. (2015) as described below. For the $i$-th point of a light curve, the selection criterion is given by the expressions:
\begin{equation}
x_\mathrm{i} - {\bar{x}_\mathrm{L}} < 0, 
\end{equation}
\begin{equation}
\frac{\left|x_\mathrm{i} - {\bar{x}_\mathrm{L}}\right|}{\sigma_\mathrm{L}} \ge N_\mathrm{1},
\end{equation}
\begin{equation}
\frac{\left|x_\mathrm{i} - {\bar{x}_\mathrm{L}} + w_\mathrm{i}\right|}{\sigma_\mathrm{L}} > N_\mathrm{2},
\end{equation}
\begin{equation}
ConM \ge N_\mathrm{3},
\end{equation} where the mean $\bar{x}_\mathrm{L}$ and deviation $\sigma_\mathrm{L}$ are the local statistics for a given segment, $w_\mathrm{i}$ is the photometric error at epoch $i$, and $ConM$ is the number of consecutive points which satisfy the equations (1--3).  The values of $N_{1,2,3}$ are taken to be 3, 2, and 3, respectively.

We used the Savitzky-Golay filter ({\it ``savgol'' package}: \citealt{doi:10.1021/ac60214a047}) to the PDCSAP light curve for detrending any long-term periodicity. {\it find\_flares()} package is used to detect the flare in the detrended light curve, which gives flare duration, starting and ending time of the flare, and equivalent duration (ED) for the detected flare. We found two flares and estimated the flare energy using the bolometric flare luminosity of $5.4\times10^{32}$ erg for $A_v=$ 6.37 (\cite{2018ApJ...861...76P}). We estimated the flare energies as $2.64\times10^{36}$ erg and $4.59\times10^{35}$ erg following the flare energy calculation method described in \citep{2018ApJ...861...76P, 2021MNRAS.500.5106G}. We summarised the properties of the flares in Table\,\ref{tess_flare_tbl}. The PDCSAP and normalized light curves (Fig.\,\ref{tess_ct4}, top left panel) show two spikes indicating the flares.

%
%

\section{Summary}
\label{summary}
\begin{enumerate}

\item The light curve of the variable object 2MASS J03435638+3209591 from IC~348 shows significant variability with RMS=0.105 mag with a short period of 4.24 hours found from the LS periodogram analysis.
\item From the $I$-band light-curve analysis of the CT2. CT3 and CT4, we detected hour-scale photometric variability using the LS periodogram method.  We find that the detected periods are varied in various observing dates, ranging from 1.5 to 3 hours.
\item We find here that TESS data gives a long-term stable period of 2.99 days for CT4 in sectors 43, as reported in previous studies.
\item From sector 43 of TESS 2-min cadence data of CT4, we detected two flare events with energies $4.59\times10^{35}$ erg and $2.64\times10^{36}$ erg, which sit in the  superflare range.
\end{enumerate}

\begin{landscape}

\begin{table}
\centering
\begin{minipage}{210mm}
\caption{Observation Log for Taurus, where ``Run-Length'' is the total duration of the observation.}
\label{log-table}
\end{minipage}
\bigskip

\begin{tabular}{cccccccc}
\hline
\textbf{Date} & \textbf{Target} & \textbf{Telescope} & \textbf{Instrument} & \textbf{FoV}          & \textbf{Run-Length} & \textbf{N$\times$Exposure} & \textbf{Seeing}   \\
              &                 &                    &                     & \textbf{(arcmin$^2$)} & \textbf{(hours)}    & \textbf{(sec)}               & \textbf{(arcsec)} \\
\hline
30/01/2009  & CT4      & 1-m ST     & 2k Wright ccd & 13$\times$13 & 4.10 & 46$\times$300                & 1.14 \\
31/01/2009  & CT2, CT3 & 1-m ST     & 2k Wright ccd & 13$\times$13 & 5.81 &  1$\times$400, 17$\times$500 & 1.22 \\
23/10/2009  & CT4      & 1-m ST     & 2k Wright ccd & 13$\times$13 & 3.19 & 32$\times$300                & 1.03 \\
08/12/2018  & CT2, CT3 & 1.3-m DFOT & ANDOR 2Kx2K   & 18$\times$18 & 5.47 &  3$\times$300, 39$\times$400 & 3.06 \\
25/12/2019  & CT2, CT3 & 2-m HCT    & 2kx2k CCD     & 10$\times$10 & 5.97 & 40$\times$200, 58$\times$180 & 1.61 \\
\hline
\end{tabular}
\end{table}

\begin{table}
\centering
\begin{minipage}{210mm}
\caption{Flare parameters of CFHT-BD-Tau 4 in TESS sector 43}
\label{tess_flare_tbl}
\end{minipage}
\bigskip

\begin{tabular}{ccccccc}
\hline
\textbf{Parameters} & \textbf{Flare start} & \textbf{Flare stop} & \textbf{E.D.}  & \textbf{Duration} & \textbf{Fractional} & \textbf{Energy} \\
                    & \textbf{(BKJD)}       & \textbf{(BKJD)}      & \textbf{(sec)} & \textbf{(hh:mm)}  & \textbf{Amplitude}  & \textbf{(erg)}  \\
\hline
Flare 1 & 2489.571444 & 2489.672842 & 4883.7 $\pm$ 70.1 & 2:26 & 1.38 & $2.64\times10^{36}$ \\
Flare 2 & 2479.849736 & 2479.874738 &  851.3 $\pm$ 37.3 & 0:36 & 0.64 & $4.59\times10^{35}$ \\
\hline
\end{tabular}
\end{table}

\end{landscape}

\begin{acknowledgments}
This research work is supported by the S N Bose National Centre for Basic Sciences under the Department of Science and Technology, Govt. of India. The authors are thankful to the Joint Time Allocation Committee (JTAC) members and the staff of the 1m-ST and 1.3-m Devasthal optical telescope operated by the Aryabhatta Research Institute of Observational Sciences (ARIES, Nainital), the HCT Time Allocation Committee (HTAC) members and the staff of the Himalayan Chandra Telescope (HCT), operated by the Indian Institute of Astrophysics (IIA, Bangalore). SG is grateful to the Department of Science and Technology (DST), Govt. of India for their Innovation in Science Pursuit for Inspired Research (INSPIRE) Fellowship scheme. This paper includes data collected with the TESS mission, obtained from the Mikulski Archive for Space Telescopes (MAST) data archive at the Space Telescope Science Institute (STScI), which is operated by the Association of Universities for Research in Astronomy, Inc., under NASA contract NAS 5-26555.
\end{acknowledgments}

\begin{furtherinformation}

\begin{orcids}
\orcid{0000-0003-3354-850X}{Samrat}{Ghosh}
\orcid{0000-0003-1457-0541}{Soumen}{Mondal}
\orcid{0009-0007-1545-854X}{Santosh}{Joshi}
\orcid{0000-0001-9367-1580}{Sneh}{Lata}
\orcid{0000-0001-7277-2577}{Rajib}{Kumbhakar}
\end{orcids}

\begin{authorcontributions}
Observational data from telescopes were collected by SG, SM, SJ and SL. The light curves and power spectrum analysis were carried out by RK, SG, and SM. The examination of the flare analysis part was performed by SG and RK. The manuscript text was prepared by SG and SM. All the authors provided input on the written draft of the manuscript, as well as the discussion and interpretation of the findings.
\end{authorcontributions}

\begin{conflictsofinterest}
The authors declare no conflict of interest.
\end{conflictsofinterest}

\end{furtherinformation}

\bibliographystyle{bullsrsl2-en}
\bibliography{S07-P08_GhoshS}

\end{document}